\documentclass[prd,amsmath,amssymb,preprintnumbers,superscriptaddress,twocolumn,10pt,nofootinbib,floatfix]{revtex4-1}

\usepackage{graphicx}

\usepackage{dcolumn}
\usepackage{bm}
\usepackage{amssymb}
\usepackage{latexsym}
\usepackage{booktabs}
\usepackage{amsmath}
\usepackage{multirow}
\usepackage{url}
\usepackage{changes}
\usepackage{soul}
\usepackage{xcolor}
\usepackage[colorlinks=true, linkcolor=red, citecolor=blue]{hyperref}
\usepackage{microtype} 
\usepackage{tikz}
\usetikzlibrary{arrows,positioning,fit}
\IfFileExists{orcidlink.sty}{\usepackage{orcidlink}}{%
  \newcommand{\orcidlink}[1]{%
    \href{https://orcid.org/##1}{%
      \includegraphics[height=1.6ex]{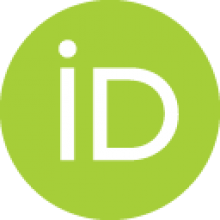}%
    }%
  }%
}

\usepackage[normalem]{ulem}
\usepackage{array}
\usepackage{enumerate}

\def\be{\begin{equation}}
\def\ee{\end{equation}}

\def\bea{\begin{eqnarray}}
\def\eea{\end{eqnarray}}
\newcommand{\bes}{\begin{equation*}}
\newcommand{\ees}{\end{equation*}}
\newcommand{\beqa}{\begin{eqnarray}}
\newcommand{\eeqa}{\end{eqnarray}}

\begin{document}

\title{Three-band dark-siren cosmology with intermediate mass black hole binaries: Synergy of Taiji, LGWA, and the Einstein Telescope}

\author{Ji-Yu Song\orcidlink{0009-0003-8111-0470}}
\affiliation{Liaoning Key Laboratory of Cosmology and Astrophysics, College of Sciences, Northeastern University, Shenyang 110819, China}

\author{Yue-Yan Dong\orcidlink{0009-0003-9108-4974}}
\affiliation{Liaoning Key Laboratory of Cosmology and Astrophysics, College of Sciences, Northeastern University, Shenyang 110819, China}

\author{Shang-Jie Jin\orcidlink{0000-0003-3697-3501}}
\affiliation{Liaoning Key Laboratory of Cosmology and Astrophysics, College of Sciences, Northeastern University, Shenyang 110819, China}

\author{Si-Ren Xiao\orcidlink{0009-0008-6742-0145}}
\affiliation{Liaoning Key Laboratory of Cosmology and Astrophysics, College of Sciences, Northeastern University, Shenyang 110819, China}

\author{Jing-Fei Zhang\orcidlink{0000-0002-3512-2804}}
\affiliation{Liaoning Key Laboratory of Cosmology and Astrophysics, College of Sciences, Northeastern University, Shenyang 110819, China}

\author{Xin Zhang\orcidlink{0000-0002-6029-1933}}\thanks{Corresponding author: \href{mailto:zhangxin@neu.edu.cn}{zhangxin@neu.edu.cn}}
\affiliation{Liaoning Key Laboratory of Cosmology and Astrophysics, College of Sciences, Northeastern University, Shenyang 110819, China}
\affiliation{MOE Key Laboratory of Data Analytics and Optimization for Smart Industry, Northeastern University, Shenyang 110819, China}
\affiliation{National Frontiers Science Center for Industrial Intelligence and Systems Optimization, Northeastern University, Shenyang 110819, China}

\begin{abstract}

Gravitational-wave (GW) dark sirens provide an independent probe of the cosmic expansion history. Their cosmological constraining power, however, depends critically on precise luminosity-distance measurements and sky localizations for cross-matching with galaxy catalogs. Multiband GW observations can track GW events across different frequency bands and thus improve both. Motivated by this, we forecast the cosmological potential of intermediate-mass black hole binaries (IMBHBs) observed by a three-band GW detector network composed of Taiji (TJ), the Lunar Gravitational-wave Antenna (LGWA), and the Einstein Telescope (ET). We simulate detectable IMBHB populations and analyze them with a hierarchical Bayesian dark-siren framework that includes galaxy-catalog completeness and redshift uncertainties. We find that the TJ-LGWA-ET network outperforms all two-detector configurations considered here. In the $\Lambda$CDM model, it constrains the Hubble constant and matter density to $\sim 0.12\%$ and $\sim 0.6\%$, respectively. In the $w$CDM model, a 4-year dark-siren sample alone constrains the dark-energy equation-of-state parameter $w$ to $\sim 2.7\%$. Adding baryon acoustic oscillation (BAO) and Type Ia supernova (SNe Ia) data improves the $w$ constraint to $\sim 2.1\%$, slightly better than that from the current CMB+BAO+SNe Ia combination. We also show that the final constraints remain sensitive to IMBHB population assumptions and galaxy-catalog limitations, which highlights the need for deep galaxy surveys with precise redshift measurements.

\end{abstract}


\maketitle

\section{Introduction}\label{sec:introduction}

The $\Lambda$CDM model, despite its success as the standard cosmological model, faces growing internal tensions as observational precision improves. Most notably, the Hubble constant ($H_0$) inferred from cosmic microwave background (CMB) data, $H_0=67.66\pm0.42~{\rm km~s^{-1}~Mpc^{-1}}$ \cite{Planck:2018vyg}, disagrees at approximately $6\sigma$ with the local distance-ladder measurement, $H_0=73.18\pm0.88~{\rm km~s^{-1}~Mpc^{-1}}$, based on HST and JWST observations \cite{Riess:2025chq}. This Hubble tension \cite{Verde:2019ivm} has not yet been resolved by any identified systematic effect. Meanwhile, the DESI baryon acoustic oscillation (BAO) data combined with CMB favor dynamical dark energy at $\sim3\sigma$ \cite{DESI:2025zgx}, sparking wide discussions \cite{Wu:2024faw,Li:2024qso,Du:2024pai,Li:2024qus,Ye:2025ark,Pang:2025lvh,Wu:2025wyk,Li:2025owk,Du:2025iow,Feng:2025mlo,Ling:2025lmw,Li:2025eqh,Li:2025dwz,Li:2025htp,Du:2025xes,Wu:2025vfs,Zhou:2025nkb,Zhang:2025dwu,Li:2025muv,Li:2025vqt,Li:2025ops,Halder:2025ytq,Pan:2025qwy,Yang:2025uyv,Cai:2025mas,Huang:2025som,Wang:2024dka,Pedrotti:2025ccw,Jiang:2024viw,Jiang:2024xnu,Giare:2024smz,Liu:2024yib,Colgain:2025nzf,Yao:2025kuz,Giare:2025ath,Yao:2025twv,Liu:2025myr,Li:2025vuh,Cheng:2025yue,Liu:2025mub,Li:2026xaz,Yin:2026gss,Du:2026cly,Feng:2026pzs}. However, the preferred phantom-crossing evolution from $w<-1$ to $w>-1$ acts to worsen the Hubble tension, and the DESI BAO and CMB datasets themselves exhibit $\sim2\sigma$ mutual inconsistency \cite{DESI:2025zgx,Ye:2025ark}. These tensions highlight the importance of constraining the Hubble constant and dark energy equation-of-state (EoS) parameters independent of both CMB and the distance ladder.

Gravitational-wave (GW) standard sirens offer a unique avenue for achieving this goal. Conventional late-universe probes such as BAO and Type Ia supernovae (SNe Ia) measure only relative distances and therefore require external calibration from CMB or the distance ladder to determine $H_0$. In contrast, GW standard sirens extract the absolute luminosity distance of the source directly from the waveform amplitude, relying solely on general relativity \cite{Schutz:1986gp}. This enables standard sirens to provide a truly independent probe for the cosmic expansion history \cite{Zhao:2010sz,Cai:2016sby,Chen:2017rfc,Wang:2018lun,Zhang:2018byx,Feeney:2018mkj,Zhang:2019ylr,Zhao:2019gyk,Zhang:2019loq,Wang:2019tto,Gray:2019ksv,Yu:2020vyy,Jin:2020hmc,Zhu:2021aat,Zhu:2021bpp,Qi:2021iic,Jin:2021pcv,Wang:2021srv,LIGOScientific:2021aug,Mastrogiovanni:2021wsd,Wu:2022dgy,Song:2022siz,Jin:2022tdf,Jin:2022qnj,Wang:2022oou,Jin:2023tou,Yu:2023ico,Jin:2023sfc,Han:2023exn,Jin:2023zhi,Li:2023gtu,Mastrogiovanni:2023emh,Mastrogiovanni:2023zbw,Feng:2024mfx,Feng:2024lzh,Han:2024sxm,Dong:2024bvw,Xiao:2024nmi,Zhu:2024qpp,Zheng:2024mbo,Perna:2024lod,Giare:2024syw,Han:2025fii,Feng:2025wbz,Song:2025ddm,Xiao:2025mcg,Dong:2025ikq,Zhang:2025yhi,Du:2025odq,Zhan:2025jqg,Su:2025zuc,Zhu:2026gsu}. See Ref.~\cite{Jin:2025dvf} for an up-to-date review of using GW standard sirens to constrain cosmological parameters. More importantly, the independent $H_0$ measurement from standard sirens can break the degeneracy between $H_0$ and the sound horizon $r_{\rm d}$ in BAO analyses, as well as the degeneracy between $H_0$ and the absolute magnitude $M_B$ in SNe Ia analyses \cite{Giare:2024syw,Zheng:2024mbo}. Consequently, combining GW standard sirens with BAO and SNe Ia can simultaneously constrain the Hubble constant and dark energy EoS parameters, opening a unique observational window into the late-universe expansion history and dark energy properties \cite{Song:2025bio}.

Nevertheless, the central challenge in employing standard sirens for cosmological inference lies in obtaining redshifts of GW sources. For ``bright sirens'' with electromagnetic (EM) counterparts, such as GW170817, precise spectroscopic redshifts can be obtained by identifying the host galaxy \cite{Schutz:1986gp,LIGOScientific:2017adf, LIGOScientific:2017vwq, LIGOScientific:2017zic}. However, the vast majority of GW events are not expected to possess detectable EM counterparts and are classified as ``dark sirens'' \cite{Schutz:1986gp,DES:2019ccw,LIGOScientific:2021aug}. For such events, the standard approach is to cross-match the luminosity distance and sky localization provided by GW observations with galaxy catalogs and to perform statistical inference on cosmological parameters within a hierarchical Bayesian framework \cite{Chen:2017rfc,Gray:2019ksv,Gray:2021sew,Gray:2023wgj,Mastrogiovanni:2023emh,Mastrogiovanni:2023zbw}. The cosmological constraining power of the dark siren method, therefore, depends critically on the precisions of luminosity distance measurements and sky localizations of GW events.

Multiband observation offers a promising route to substantially enhancing dark siren cosmology. For a given compact binary system, a multiband detector network can continuously track its GW signal across different evolutionary stages, thereby accumulating higher signal-to-noise ratios (SNRs), longer-baseline information, and orbital modulation effects that significantly improve luminosity distance measurements and sky localizations \cite{Sesana:2016ljz,Vitale:2016rfr, Sesana:2017vsj,Isoyama:2018rjb,Jani:2019ffg,Carson:2019kkh,Liu:2020nwz,Datta:2020vcj,Zhang:2021pwe,Nakano:2021bbw,Yang:2021qge,Muttoni:2021veo,Liu:2021dcr,Zhu:2021bpp,Kang:2022nmz,Klein:2022rbf,Seymour:2022teq,Baker:2022eiz,Zhao:2023ilw, Dong:2024bvw}. Previous studies have demonstrated that multiband observations can significantly enhance the constraining power of dark sirens on cosmological parameters \cite{Seymour:2022teq,Dong:2024bvw}. Among all potential multiband GW sources, intermediate-mass black hole binaries (IMBHBs) are particularly compelling. IMBHBs occupy a critical mass range between stellar-mass binary black holes and supermassive black hole binaries; their formation mechanisms, population distributions, and potential as the seeds of supermassive black holes make them a key topic in astrophysical and cosmological research \cite{Rosswog:2008ie, Farrell:2009uxm, Clausen:2010hf, Haas:2012bk, Pasham2014, MacLeod:2015jma,Mezcua:2017npy, Kiziltan2017, Kaaret:2017tcn,Li:2025pyo}. From the standpoint of GW detection, the GW signals from IMBHBs enter the millihertz band during the early inspiral, subsequently sweep through the decihertz band, and ultimately merge in the hectohertz band, making them natural candidates for joint observation across three frequency bands. Recent detections of high-mass binary black hole mergers confirm that this mass range harbors a viable source population that must be seriously considered in future GW astronomy and cosmology \cite{LIGOScientific:2020iuh,LIGOScientific:2025rsn}.

Realizing three-band observations of IMBHBs requires a GW detector network that spans the millihertz, decihertz, and hectohertz bands. In this work, we consider a representative network composed of Taiji (TJ) \cite{Hu:2017mde}, the Lunar Gravitational-wave Antenna (LGWA) \cite{LGWA:2020mma}, and the Einstein Telescope (ET) \cite{Punturo:2010zz,ET:2025xjr}. TJ probes the early inspiral in the millihertz band, ET captures the late inspiral, merger, and ringdown in the hectohertz band, and LGWA bridges the intermediate decihertz band. Together, this TJ-LGWA-ET network provides continuous frequency coverage from the millihertz to the hectohertz band, allowing us to track IMBHB signals across their full evolution and substantially improving luminosity-distance measurements and sky localizations. Based on this network, we forecast the cosmological constraining power of the three-band IMBHB dark sirens. We simulate the detectable IMBHB population, evaluate the improvements in luminosity distance and sky localization enabled by three-band observations, and propagate them into cosmological parameter constraints within a dark-siren statistical framework. We also explore how GW population models and galaxy-catalog properties affect the final constraints.

This paper is organized as follows. Sect.~\ref{sec:method} describes the detector network configuration, source population models, detection simulation methodology, and the dark siren Bayesian analysis framework. Sect.~\ref{sec:results} presents the cosmological constraint results under different network configurations and discusses the influences of GW population models and galaxy catalogs. Finally, we summarize the main conclusions and outline future research directions.


\section{Method}\label{sec:method}

\subsection{Three-band GW network for IMBHBs}

\begin{figure*}[tp]
    \centering
    \includegraphics[width=0.8\linewidth]{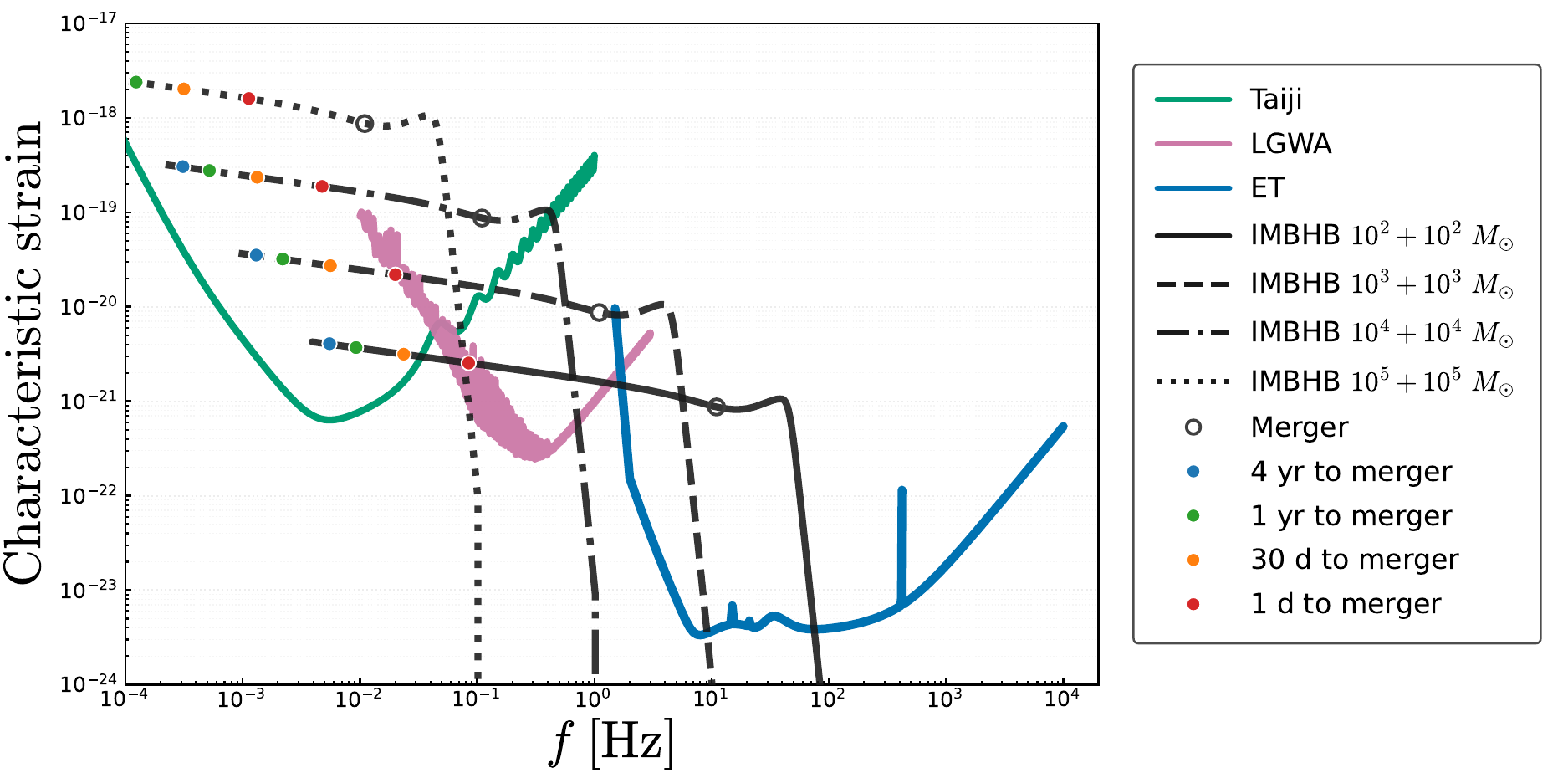}
    \caption{Comparison between multi-band GW detector sensitivities and the frequency-domain detectability of IMBHBs. The horizontal axis shows the frequency and the vertical axis shows the characteristic strain, both on logarithmic scales. The colored solid lines display the noise characteristic-strain curves of three detectors: TJ (green), LGWA (pink), and ET (blue). The four black curves with different line styles represent the signal tracks of equal-mass IMBHBs ($10^2\!+\!10^2$, $10^3\!+\!10^3$, $10^4\!+\!10^4$, and $10^5\!+\!10^5\,M_\odot$) at $z=1$, simulated using the \textsc{IMRPhenomD} waveform and converted to characteristic strain via $h_c = 2f|\tilde{h}_+|$. Open circles along each black track mark the approximate beginnings of the merger phase. The blue, green, orange, and red filled markers indicate the signal locations at $4$ years, $1$ year, $30$ days, and $1$ day before the merger.}
    \label{fig:fh}
\end{figure*}

In this subsection, we specify the detector network adopted in this work and clarify why IMBHBs enable three-band observations. We consider a GW detector network composed of the space-based observatory TJ, the moon-based detector LGWA, and the 3G ground-based detector ET. This network provides continuous sensitivity from the millihertz to the hectohertz band, as shown in Fig.~\ref{fig:fh}. For each instrument, we adopt a fiducial strain noise power spectral density (PSD) and present the corresponding characteristic strain noise curve. We adopt the latest Taiji design sensitivity for TJ, the ET-D design sensitivity~\cite{Punturo:2010zz,Hild:2010id} for ET, and the white-paper sensitivity~\cite{LGWA:2020mma} for LGWA. We adopt this fiducial ET-D curve for consistency with previous multiband studies, while noting that it is an early design and may be comparatively optimistic at low frequency relative to more recent ET sensitivity curves. LGWA has two proposed configurations, with either niobium or silicon as the proof-mass and suspension material \cite{Ajith:2024mie}. In this work, we adopt the silicon-based configuration, which offers better sensitivity.

We model IMBHB signals using the \textsc{IMRPhenomD} waveform model~\cite{Husa:2015iqa,Khan:2015jqa}. The frequency-domain waveform vector for a network of $N$ interferometers is given by~\cite{Wen:2010cr,Zhao:2017cbb}
\begin{equation}
\bm{\tilde{h}}(f) = e^{-i\bm{\Phi}(f)} \, \bm{\hat{h}}(f),
\end{equation}
where $\bm{\Phi}(f)$ is an $N\times N$ diagonal matrix describing the propagation-delay phase, with elements
\begin{equation}
\Phi_{kl}(f)=2\pi f\,\delta_{kl}\,\frac{\bm{n}\cdot\bm{r}_k}{c},
\end{equation}
where $\bm{n}$ is the GW propagation direction and $\bm{r}_k$ is the position vector of the $k$-th interferometer. The waveform vector $\bm{\hat{h}}(f)$ is defined as
\begin{equation}
\bm{\hat{h}}(f) = \left[\tilde{h}_1(f), \tilde{h}_2(f), \cdots, \tilde{h}_N(f)\right],
\end{equation}
with each component given by
\begin{equation}
\tilde{h}_k(f)=h_{+}(f)F_{+,k}(f)+h_{\times}(f)F_{\times,k}(f),
\end{equation}
where $h_{+}(f)$ and $h_{\times}(f)$ are the two GW polarizations generated with \textsc{IMRPhenomD}, and $F_{+,k}(f)$ and $F_{\times,k}(f)$ are the antenna-pattern functions of the $k$-th interferometer. We take the antenna-pattern functions for TJ from Ref.~\cite{Ruan:2020smc} and those for ET from Ref.~\cite{Punturo:2010zz}. For LGWA, we use the public code \texttt{GWFish}~\cite{Dupletsa:2022scg} to model the detector response \cite{Yan:2024jio}.

Fig.~\ref{fig:fh} shows characteristic-strain tracks for equal-mass binaries with source-frame component masses $(10^2,10^3,10^4,10^5)\,M_\odot$, evaluated at a reference redshift $z_{\rm ref}=1$. Detector-frame masses are
\begin{equation}
m_i^{\rm det}=(1+z_{\rm ref})\,m_i^{\rm src},
\end{equation}
and are converted to geometric units via $m_i=m_i^{\rm det}T_\odot$, with $T_\odot\equiv GM_\odot/c^3$. The chirp mass is
\begin{equation}
\mathcal{M}_c=\frac{(m_1m_2)^{3/5}}{(m_1+m_2)^{1/5}}.
\end{equation}
The colored markers (4 yr, 1 yr, 30 d, 1 d before coalescence) are calculated by the Newtonian time-to-coalescence relation \cite{Cutler:1994ys, Sathyaprakash:2009xs},
\begin{equation}
\tau=\frac{5}{256}\,\mathcal{M}_c^{-5/3}(\pi f)^{-8/3},
\end{equation}
which gives
\begin{equation}
f(\tau)=\frac{1}{8\pi\mathcal{M}_c}\left(\frac{5\mathcal{M}_c}{\tau}\right)^{3/8}.
\end{equation}
For each $\tau$, the plotted node is $\bigl[f(\tau),\,h_c(f(\tau))\bigr]$, with $h_c(f)=2f\,|\tilde h_+(f)|$, and $h_c(f(\tau))$ obtained by interpolation along the simulated track.

The open circles denote the adopted boundaries for the onset of mergers. The merger frequency is approximated by the Schwarzschild innermost stable circular orbit frequency \cite{Hanna:2008um},
\begin{equation}
f_{\rm merger}=\frac{1}{\pi\,6^{3/2}\,M},\qquad M\equiv m_1+m_2.
\end{equation}
These relations are used only to place the illustrative stage markers in Fig.~\ref{fig:fh} (the time-to-coalescence dots from $\tau$ and $f(\tau)$, and the merger-onset circles from $f_{\rm merger}$); they do not enter the SNR, Fisher-matrix, or cosmological calculations, which use the full \textsc{IMRPhenomD} waveform over the frequency range set by each detector's sensitivity band and the waveform validity. Across this mass range, the TJ+LGWA+ET network enables gapless multiband tracking: signals enter in the mHz band (TJ) during early inspiral and proceed to merger in either LGWA or ET sensitive bands.


\subsection{IMBHB populations and detectable samples}\label{sec:pop}

To model the cosmic merger history of IMBHBs, we adopt the phenomenological framework of Ref.~\cite{Fragione:2022ams}, which parameterizes the merger-rate density as
\begin{equation}
    R(z, M_1, q)
    = K\,N(\mu_z, \sigma_{\rm R})\,M_1^{-\alpha}\,q^{-\beta},
\end{equation}
where $K$ is a normalization constant (in units of $\mathrm{Gpc^{-3}\,yr^{-1}}$), $N(\mu_z, \sigma_{\rm R})$ describes the redshift evolution of the merger rate and is modeled as a normal distribution with the mean value $\mu_z$ and the 1$\sigma$ width $\sigma_{\rm R}$, $M_1$ is the primary mass, and $q\equiv m_2/m_1\leq1$ is the mass ratio. The power-law factors in $M_1$ and $q$ provide a flexible low-dimensional description of the IMBHB mass and mass-ratio distributions, while the redshift term captures the broad uncertainty in the cosmic evolution of IMBH mergers. This parameterization is intended to encompass a variety of formation channels discussed in the literature, such as those in dense stellar clusters and galactic nuclei~\cite{Fragione:2017blf,Fragione:2018lvy,Rasskazov:2019tgb,Fragione:2022egh,Fragione:2021nhb}. The corresponding event number per observer time is
\begin{equation}
    \frac{{\rm d}^3\dot{N}}{{\rm d}z\,{\rm d}M_1\,{\rm d}q}
    =
    \frac{R(z,M_1,q)}{1+z}\,\frac{{\rm d}V_c}{{\rm d}z},
\end{equation}
where ${\rm d}V_c/{\rm d}z$ is the differential comoving volume.

Unless otherwise stated, we follow the baseline population choices of Ref.~\cite{Fragione:2022ams}: $(\alpha,\beta)=(1,1)$ and $\sigma_{\rm R}=1$, and draw binaries from $M_1\in[10^2,10^5]\,M_\odot$ and $q\in[0.01,1]$. We further impose $m_2>10\,M_\odot$ and discard systems violating this cut to suppress an atypical accumulation of extremely small companions near the low-$q$ boundary. For the redshift evolution, we consider two representative peak models, denoted as $z2$ and $z5$, with $\mu_z=2$ and $\mu_z=5$, respectively, representing merger histories peaking at relatively low and high redshift, respectively. We sample redshifts over a broad interval, $z\in[10^{-8},10]$. For the overall rate normalization, we set $K=10~\mathrm{Gpc^{-3}\,yr^{-1}}$ in the baseline implementation, corresponding to an optimistic choice within the range discussed in Ref.~\cite{Fragione:2022ams}. We assume isotropic sky locations and binary orientations. We note that high-total-mass BBH mergers forming intermediate-mass remnants have now been observed at low-to-moderate redshift, including GW190521 (remnant mass $\sim 150\,M_\odot$, $z\simeq0.8$)~\cite{LIGOScientific:2020iuh} and GW231123 (source-frame total mass $190$--$265\,M_\odot$, $z\simeq0.4$)~\cite{LIGOScientific:2025rsn}. These detections confirm that mergers near the low-mass end of our IMBHB population do occur, albeit at a low rate. For GW231123-like events, the single-event rate estimate is $0.08^{+0.19}_{-0.07}~\mathrm{Gpc^{-3}\,yr^{-1}}$~\cite{LIGOScientific:2025rsn}, well below the overall BBH merger rate inferred from current catalogs, $\sim 16$--$61~\mathrm{Gpc^{-3}\,yr^{-1}}$~\cite{LIGOScientific:2025pvj}. This comparison, however, constrains only the low-redshift, high-total-mass tail currently accessible to ground-based detectors, and does not determine the merger history at the higher redshifts targeted by our forecasts~\cite{Jani:2019ffg}. 
Our $z2$ and $z5$ models span a representative range of this uncertain evolution. 
{With the baseline normalization $K=10~\mathrm{Gpc^{-3}\,yr^{-1}}$, the implied local merger-rate densities are $R(z=0)\simeq17~\mathrm{Gpc^{-3}\,yr^{-1}}$ for the $z2$ model and $R(z=0)\simeq5\times10^{-4}~\mathrm{Gpc^{-3}\,yr^{-1}}$ for the $z5$ model, where $R(z=0)=K\,N(z=0,\mu_z,\sigma_{\rm R})\iint M_1^{-1}q^{-1}\,dM_1\,dq$ over the adopted mass and mass-ratio ranges. The $z2$ value exceeds the LVK-inferred rate for IMBHBs by about two orders of magnitude, while the $z5$ value lies about two orders of magnitude below it, so the two models bracket the LVK value. The $z2$ and $z5$ models are therefore the optimistic upper-end and conservative lower-end cases, respectively.}

For each simulated source, we evaluate the matched-filter SNR in each interferometer channel, and for a detection network composed of $N$ interferometers, the total SNR is given by
\begin{equation}
    \rho = \sqrt{ \sum_{i=1}^{N}\left( \tilde{h}_i | \tilde{h}_i \right) },
\end{equation}
where the inner product is defined as
\begin{equation}
    \left( \tilde{h} | \tilde{h} \right) = 4 \int_{f_{\rm min}}^{f_{\rm max}} \frac{\tilde{h}(f) \tilde{h}^*(f)}{S_{{\rm n}}(f)} \, df,
\end{equation}
where $S_{\rm n}(f)$ is the one-sided noise PSD. The integration limits $f_{\min}$ and $f_{\max}$ are set by each detector's sensitive band ($[10^{-5},1]\,$Hz for TJ, $[10^{-3},4]\,$Hz for LGWA, and $[1,10^4]\,$Hz for ET), intersected with the band over which the \textsc{IMRPhenomD} waveform is defined. An event is classified as detectable if $\rho_{\rm net}\geq 8$. In this way, we construct the detectable samples for each detector network and compare their redshift distributions.

\begin{figure}[htbp]
    \centering
    \includegraphics[width=1\linewidth]{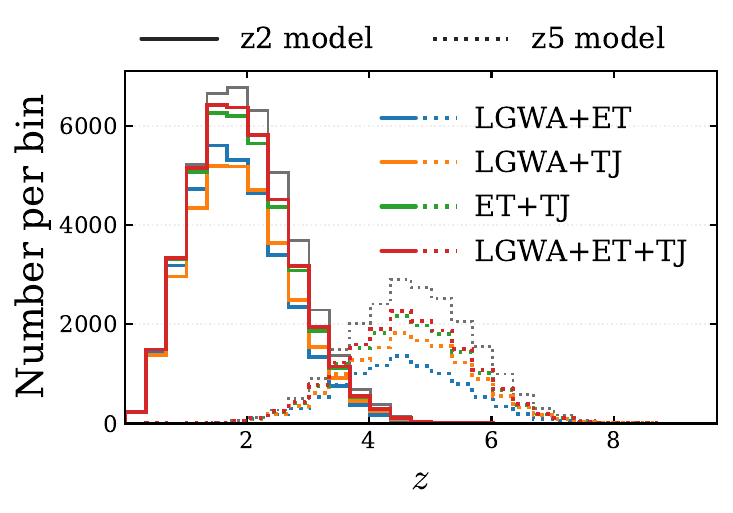}
    \caption{Redshift distributions of detectable events for different detector networks under the detection threshold of $\mathrm{SNR}\ge 8$ during 1-year observation. The colored step histograms show the network-selected samples for LGWA+ET, LGWA+TJ, ET+TJ, and LGWA+ET+TJ, where the solid lines correspond to the $z2$ population model and the dotted lines correspond to the $z5$ population model; colors distinguish detector networks. The gray solid and gray dotted curves denote the total IMBHB distributions for the $z2$ and $z5$ models, respectively.}
    \label{fig:Nz}
\end{figure}

\begin{figure}[htbp]
    \centering
    \includegraphics[width=1\linewidth]{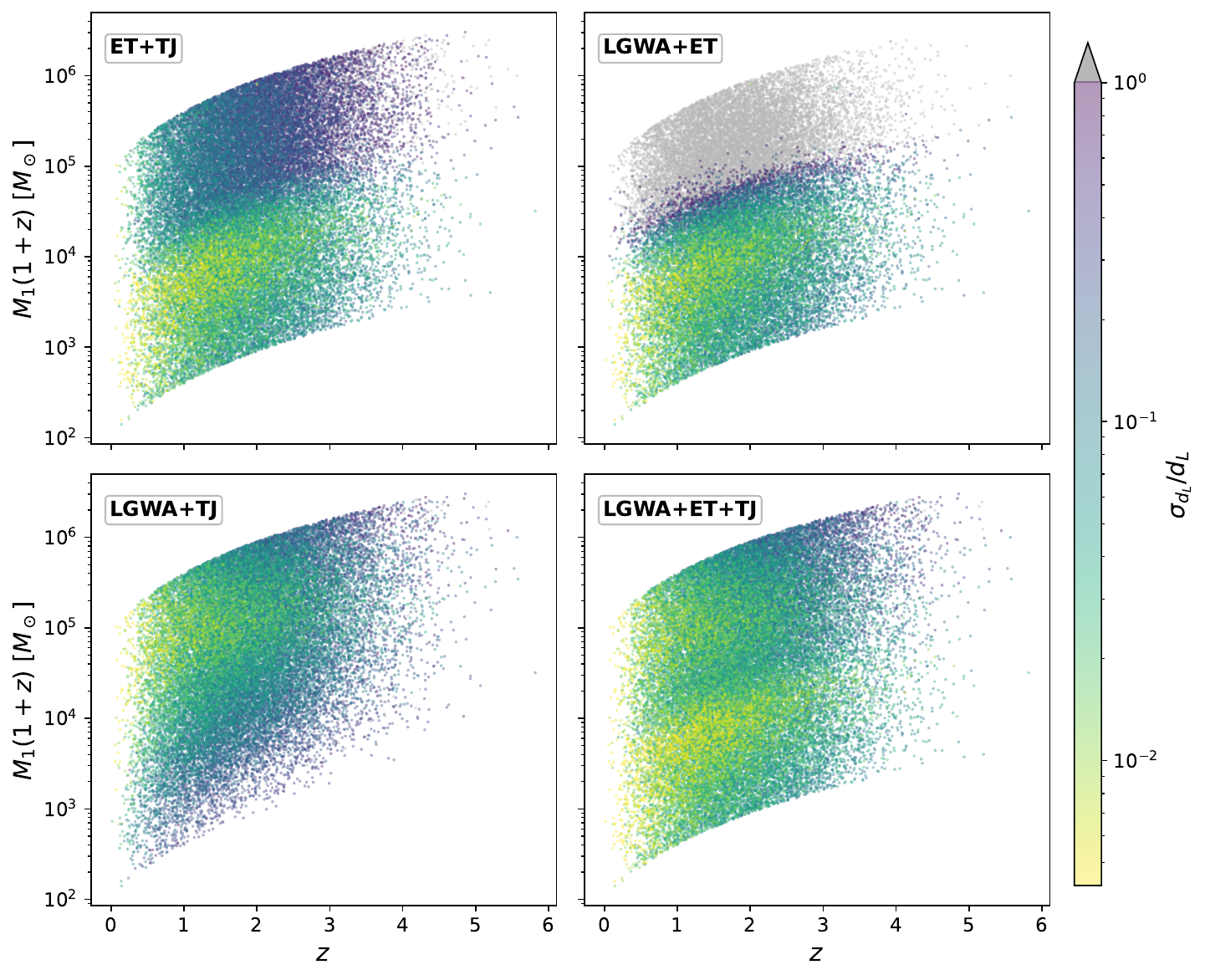}
    \caption{Detector-frame primary mass $M_1(1+z)$ versus redshift $z$ for the detected IMBHB population ($z2$ model, 1-year observation, $\rho_{\rm net}\geq8$), shown separately for the four detector networks. Each point is one event, color-coded by its relative luminosity-distance uncertainty $\sigma_{d_{\rm L}}/d_{\rm L}$; gray points (above the colorbar range) have $\sigma_{d_{\rm L}}/d_{\rm L}>1$, i.e., an essentially unconstrained distance. The rising lower envelope reflects the selection effect, while the color pattern shows which systems each network measures best.}
    \label{fig:m1z}
\end{figure}

Fig.~\ref{fig:Nz} summarizes the redshift distributions of detectable events for different detector networks. For the $z2$ model (solid curves), detectable events mainly cluster around $z\sim1.5$--$2.5$, and the three-band network LGWA+ET+TJ yields the largest number of detections. For the $z5$ case (dotted curves), the peak shifts to $z\sim4$--$5$ with a broader high-$z$ tail: LGWA+ET+TJ still performs best and reaches the highest redshift, ET+TJ is the next most effective, while LGWA+TJ and especially LGWA+ET drop more rapidly at high $z$. To illustrate what the detected population looks like, Fig.~\ref{fig:m1z} shows the detector-frame primary mass $M_1(1+z)$ versus redshift for the detectable events ($\rho_{\rm net}\geq8$) of each detector network, color-coded by the relative luminosity-distance error $\sigma_{d_{\rm L}}/d_{\rm L}$. The detectable sample spans $M_1(1+z)\sim10^2$--$10^6\,M_\odot$; the resulting measurement precision and its dependence on the source parameters and the detector network are discussed in Sec.~\ref{sec:fisher}.

\subsection{GW parameter estimation}
\label{sec:fisher}

For each detected IMBHB event, we forecast the source-parameter measurement uncertainties arising from instrument noise using the Fisher information matrix (FIM) \cite{Cutler:1994ys}. For a detector network, FIM is
\begin{equation}
\Gamma_{ij}
=
\sum_I^{N}
\left(
\frac{\partial \tilde{h}_I}{\partial \lambda_i}
\Bigg|
\frac{\partial \tilde{h}_I}{\partial \lambda_j}
\right)_I,
\end{equation}
where $\lambda_{i/j}$ is the $i/j$-th parameter. We adopt the parameter vector
\begin{equation}
\boldsymbol{\lambda}
=
\left\{
\mathcal{M}_c,\,
q,\,
d_{\rm L},\,
\iota,\,
\delta,\,
\alpha,\,
\psi,\,
\phi_c,\,
t_c
\right\},
\end{equation}
where $\mathcal{M}_c$ is the detector-frame chirp mass, $q\equiv m_2/m_1\le 1$ the mass ratio, $d_{\rm L}$ the luminosity distance, $\iota$ the inclination angle, $(\alpha,\delta)$ the source right ascension and declination, $\psi$ the polarization angle, and $(\phi_c,t_c)$ the coalescence phase and time.

The covariance matrix is approximated by
\begin{equation}
\Sigma \simeq \Gamma^{-1},
\end{equation}
so that the $1\sigma$ uncertainty of $\lambda_i$ is $\sigma_{\lambda_i}=\sqrt{\Sigma_{ii}}$.

The instrumental distance uncertainty is obtained directly from the covariance matrix,
\begin{equation}
\left(\frac{\sigma_{d_{\rm L}}}{d_{\rm L}}\right)_{\rm inst}
=
\frac{\sqrt{\Sigma_{d_{\rm L} d_{\rm L}}}}{d_{\rm L}}.
\end{equation}
For cosmological forecasts, we further combine this term with the contributions from peculiar velocity and weak lensing,
\begin{equation}\label{eq:dl_err}
\left(\frac{\sigma_{d_{\rm L}}}{d_{\rm L}}\right)^2
=
\left(\frac{\sigma_{d_{\rm L}}}{d_{\rm L}}\right)_{\rm inst}^2
+
\left(\frac{\sigma_{d_{\rm L}}}{d_{\rm L}}\right)_{\rm pv}^2
+
\left(\frac{\sigma_{d_{\rm L}}}{d_{\rm L}}\right)_{\rm lens}^2.
\end{equation}
The peculiar-velocity contribution represents the uncertainty in the source redshift, $\sigma_z=(1+z)\sigma_v/c$, induced by its peculiar motion and propagated into the luminosity-distance error through the $d_{\rm L}$--$z$ relation:
\begin{equation}
\left(\frac{\sigma_{d_{\rm L}}}{d_{\rm L}}\right)_{\rm pv}
=
\left[1+\frac{c\,(1+z)^2}{H(z)\,d_{\rm L}(z)}\right]
\frac{\sigma_v}{c},
\end{equation}
where $\sigma_v=500~{\rm km\,s^{-1}}$ \cite{Kocsis:2005vv}, and the weak lensing error is modeled as \cite{Tamanini:2016zlh}
\begin{equation}
\left(\frac{\sigma_{d_{\rm L}}}{d_{\rm L}}\right)_{\rm lens}
=
0.066
\left[
\frac{1-(1+z)^{-0.25}}{0.25}
\right]^{1.8}.
\end{equation}

The sky-localization uncertainty is obtained from the covariance submatrix of $(\delta,\alpha)$,
\begin{equation}\label{eq:sky_localization}
\Sigma_{\rm sky}
=
\begin{pmatrix}
\Sigma_{\delta\delta} & \Sigma_{\delta\alpha}\\
\Sigma_{\alpha\delta} & \Sigma_{\alpha\alpha}
\end{pmatrix},
\end{equation}
which gives the $1\sigma$ solid-angle uncertainty
\begin{equation}
\Delta\Omega_{1\sigma}
=
2\pi |\cos\delta|
\sqrt{
\Sigma_{\delta\delta}\Sigma_{\alpha\alpha}
-
\Sigma_{\delta\alpha}^2
}.
\end{equation}
Assuming a Gaussian posterior, the corresponding $90\%$ credible localization area is
\begin{equation}
\Delta\Omega_{90\%}
=
4.605\,\Delta\Omega_{1\sigma}.
\end{equation}

\begin{figure}[htbp]
    \centering
    \includegraphics[width=1\linewidth]{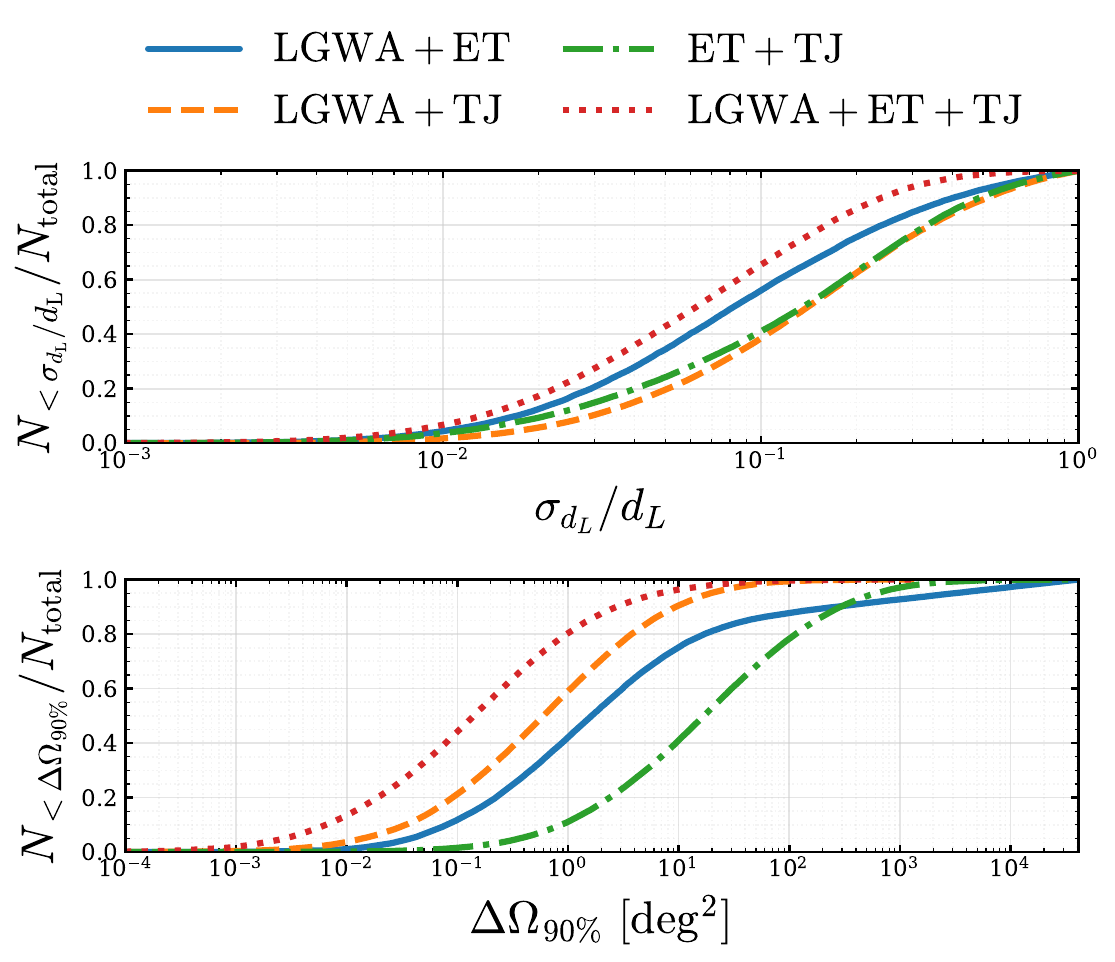}
    \caption{Cumulative distribution functions (CDFs) of the relative luminosity distance error $\sigma_{d_{\rm L}}/d_{\rm L}$ and the 90\% sky localization error $\Delta \Omega_{90\%}$ for different detector networks. The top panel shows the CDF of the fractional luminosity-distance uncertainty, where the y-axis value at a given x-axis value $X$ in the CDF curve represents the proportion of GW events with $\sigma_{d_{\rm L}}/d_{\rm L}$ less than $X$ relative to the total number of events. The bottom panel shows the CDF of the 90\% credible sky-localization area, $\Delta\Omega_{90\%}$ deg$^2$, following analogously.}
    \label{fig:CDF}
\end{figure}

Fig.~\ref{fig:CDF} summarizes CDFs of the estimated measurement errors of luminosity distances and sky localizations for different GW detector networks. In the upper panel, LGWA+ET+TJ yields the best luminosity-distance precision, with its CDF systematically shifted to smaller $\sigma_{d_{\rm L}}/d_{\rm L}$. LGWA+ET performs next best, followed by ET+TJ, while LGWA+TJ is comparatively the least constraining. The lower panel shows an even stronger hierarchy in sky localization. The LGWA+ET+TJ network again performs best, followed by LGWA+TJ and LGWA+ET, while ET+TJ gives the broadest localization areas for most events. Since dark-siren cosmology depends on the three-dimensional GW localization volume, these improvements in both $\sigma_{d_{\rm L}}/d_{\rm L}$ and $\Delta\Omega_{90\%}$ are the key ingredients behind the tighter cosmological constraints.

Fig.~\ref{fig:m1z} further clarifies which sources drive these constraints. The smallest $\sigma_{d_{\rm L}}/d_{\rm L}$ is always attained at low redshift, but the optimal mass depends on the frequency coverage: networks including ET reach their best precision for detector-frame masses $M_1(1+z)\sim10^3$--$10^4\,M_\odot$ (which merge in the hectohertz band), whereas LGWA+TJ performs best for $M_1(1+z)\sim10^5\,M_\odot$ (the millihertz--decihertz band). More importantly, a high SNR alone does not guarantee a usable distance: for the LGWA+ET network, $\sim36\%$ of detected events (gray points in Fig.~\ref{fig:m1z}) have $\sigma_{d_{\rm L}}/d_{\rm L}>1$, i.e., an essentially unconstrained distance, because neither detector provides the long-timescale antenna-pattern modulation needed to break the distance--inclination degeneracy. Adding TJ, whose heliocentric orbital motion modulates the detector response over the long inspiral, breaks this degeneracy and reduces the unconstrained fraction to $5.9\%$ (ET+TJ), $0.5\%$ (LGWA+TJ), and only $0.2\%$ for the full LGWA+ET+TJ network. The three-band network therefore not only detects the most events but also delivers a usable luminosity distance for almost all of them, which underlies its superior cosmological performance.

\subsection{Dark-siren analysis}

\begin{figure*}[tp]
\centering
\small
\begin{tikzpicture}[
    >=latex,
    node distance=1.2cm and 1.2cm,
    every node/.style={align=center},
    block/.style={
        draw,
        rounded corners,
        thick,
        inner sep=6pt,
        text width=0.28\textwidth
    },
    blocksmall/.style={
        draw,
        rounded corners,
        thick,
        inner sep=6pt,
        text width=0.24\textwidth
    },
    line/.style={->, thick}
    ]

\node[blocksmall] (gw) {
{\bf GW data}\\[2pt]
$\rho>\rho_{\rm th}$ \\[4pt]
$d_{{\rm L},i},\ \sigma_{d_{{\rm L}},i}$ \\[2pt]
$\Delta\Omega_{i},\ \Sigma_{{\rm sky},i}$
};

\node[block, right=of gw] (gal) {
{\bf Galaxy catalog}\\[4pt]
$p_{\rm com}(z)=
\dfrac{\int_{M_{\rm min}^{\rm obs}}^{M_{\rm th}(z)} dM\,\Phi(M)}
{\int_{M_{\rm min}^{\rm obs}}^{M_{\rm max}^{\rm obs}} dM\,\Phi(M)}$
\\[6pt]
$p_{\rm miss}(z)=1-p_{\rm com}(z)$
\\[6pt]
$\begin{aligned}
N_{{\rm exp},i} &= \mathrm{round}\!\bigg[ \Delta\Omega_i \phi_*^{\rm obs} \\
&\quad \times \int dz\, \dfrac{dV_c}{dz\,d\Omega} \, p_{\rm com}(z) \bigg]
\end{aligned}$
};

\node[blocksmall, right=of gal] (sel) {
{\bf Selection}\\[4pt]
$\begin{aligned}
p_{\rm det}(d_{\rm L}) &= \dfrac{1}{N_{\rm samp}} \\
&\quad \times \sum_{s=1}^{N_{\rm samp}} \mathcal{H}(\rho_s-\rho_{\rm th})
\end{aligned}$
};

\node[block, below=of gal] (like) {
{\bf Event likelihood}\\[4pt]
$\begin{aligned}
\Lambda_i(z\mid\Theta) &= \dfrac{R(z)}{1+z} \bigg[ N_{{\rm exp},i}p_{{\rm cat},i}(z)+ \\
&\quad  \Delta\Omega_i\phi_*^{\rm obs}\times \\
&\quad \dfrac{dV_c}{dz\,d\Omega}(z\mid\Theta) p_{\rm miss}(z) \bigg]
\end{aligned}$
\\[8pt]
$\begin{aligned}
&p(d_i\mid\Theta) = \\
&\dfrac{\int dz\,p(d_{{\rm L},i}\mid z,\Theta)\Lambda_i(z\mid\Theta)}
{\int dz\,p_{\rm det}[d_{\rm L}(z,\Theta)]\Lambda_i(z\mid\Theta)}
\end{aligned}$
};

\node[block] (host) at (gw |- like) {
{\bf Candidate hosts}\\[4pt]
$\begin{aligned}
w_{ij} &\propto \exp\!\bigg[ -\dfrac{1}{2} (\mathbf{x}_{ij}-\mathbf{x}^{\rm inj}_i)^{\rm T} \\
&\quad \times \Sigma_{{\rm sky},i}^{-1} (\mathbf{x}_{ij}-\mathbf{x}^{\rm inj}_i) \bigg]
\end{aligned}$
\\[6pt]
$\begin{aligned}
p_{{\rm cat},i}(z) &= \sum_{j=1}^{N_{{\rm exp},i}} w_{ij}\, \\
&\quad \times \mathcal{N}\!\left[z;z_{ij},\sigma_z(1+z_{ij})\right]
\end{aligned}$
};

\node[blocksmall] (post) at (sel |- like) {
{\bf Posterior}\\[4pt]
$\begin{aligned}
p(\Theta\mid\{d_i\}) &\propto \pi(\Theta) \\
&\quad \times \prod_{i=1}^{N_{\rm GW}}p(d_i\mid\Theta)
\end{aligned}$
};

\draw[line] (gw) -- (host);
\draw[line] (gw) -- (like);
\draw[line] (gal) -- (host);
\draw[line] (gal) -- (like);
\draw[line] (host) -- (like);
\draw[line] (sel) -- (like);
\draw[line] (like) -- (post);

\end{tikzpicture}
\caption{Schematic flowchart of the dark-siren analysis. The mock GW sample provides the distance and sky-localization information, which are combined with the mock galaxy catalog and the selection function to construct the event likelihood and to obtain the joint posterior for the cosmological parameters.}
\label{fig:flowchart_dark_siren}
\end{figure*}
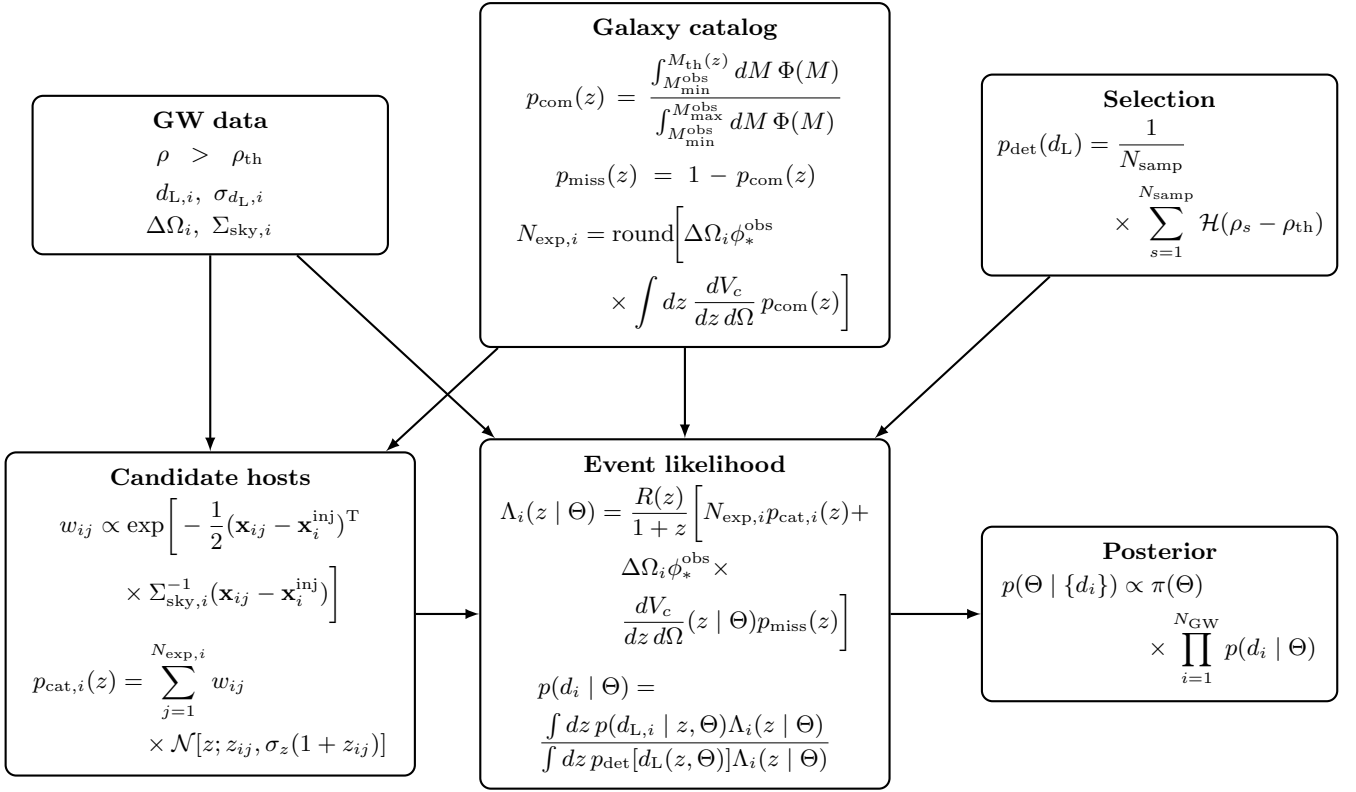

In this subsection, we present the dark-siren analysis framework, as illustrated in Fig.~\ref{fig:flowchart_dark_siren}.

\subsubsection{Bayesian framework}

We infer the cosmological parameters $\Theta$ from a set of detected dark-siren events $\{d_i\}$ according to
\begin{equation}
p(\Theta \mid \{d_i\}) \propto \pi(\Theta)\prod_{i=1}^{N_{\rm GW}} p(d_i \mid \Theta),
\end{equation}
where $\pi(\Theta)$ is the prior on the cosmological parameters and $N_{\rm GW}$ is the number of GW events included in the analysis.

For each event, the likelihood is written as
\begin{equation}\label{eq:likelihood}
p(d_i \mid \Theta)=
\frac{\int {\rm d}z \,
p(d_{{\rm L},i}\mid z,\Theta)\,
\Lambda_i(z\mid\Theta)}
{\int {\rm d}z \,
p_{\rm det}[d_{\rm L}(z,\Theta)]\,
\Lambda_i(z\mid\Theta)},
\end{equation}
where
\begin{equation}
p(d_{{\rm L},i}\mid z,\Theta)=
\frac{1}{\sqrt{2\pi}\sigma_{d_{{\rm L}},i}}
\exp\!\left[
-\frac{\bigl(d_{\rm L}(z,\Theta)-d_{{\rm L},i}\bigr)^2}{2\sigma_{d_{{\rm L}},i}^2}
\right].
\end{equation}
Here $d_{{\rm L},i}$ and $\sigma_{d_{{\rm L}},i}$ are the injected (fiducial) luminosity distance and its $1\sigma$ uncertainty for the $i$-th event, with $\sigma_{d_{{\rm L}},i}$ obtained from Eq.~\eqref{eq:dl_err}, and $d_{\rm L}(z,\Theta)$ is the distance-redshift relation predicted by the assumed cosmological model. Following standard Fisher-forecast practice, we do not add a random noise realization to each event, so the analysis returns the expected, realization-averaged sensitivity, with the measurement uncertainty carried by $\sigma_{d_{{\rm L}},i}$.

The redshift prior entering Eq.~\eqref{eq:likelihood} is
\begin{equation}\label{eq:Lambda_i}
  \begin{aligned}
    \Lambda_i(z\mid\Theta)
    =
    &\frac{R(z)}{1+z}
    \bigg[
    N_{{\rm exp},i}\,p_{{\rm cat},i}(z)
    +\\
    &\Delta\Omega_{i} \,\phi_*^{\rm obs}\,
    \frac{dV_c}{dz\,d\Omega}(z\mid\Theta)\,
    p_{\rm miss}(z)
    \bigg],
  \end{aligned}
\end{equation}
where $(1+z)^{-1}$ converts the source-frame merger rate into the detector frame. For the simulated IMBHB population considered here, as detailed in Subsect.~\ref{sec:pop}, we adopt a Gaussian merger-rate evolution,
\begin{equation}
R(z)\propto \exp\!\left[-\frac{(z-\mu_z)^2}{2\sigma_{\rm R}^2}\right],
\end{equation}
consistent with the injection model used to generate the mock GW catalog.

\subsubsection{Mock galaxy catalog and catalog completeness}

For each GW event, we construct a mock galaxy catalog within the GW sky-localization region to model the redshift prior of candidate host galaxies. Because the observable galaxy sample is flux limited, we first quantify the catalog completeness as a function of redshift.

We model the galaxy luminosity distribution with a Schechter function,
\begin{equation}
  \begin{aligned}
    \Phi(M)
      =
      &0.4\ln 10 \,\phi_*^{\rm obs}\,
      10^{0.4(\alpha+1)(M_*^{\rm obs}-M)}\times\\
      &\exp\!\left[-10^{0.4(M_*^{\rm obs}-M)}\right].
  \end{aligned}
\end{equation}
In this work, we adopt $M_*=-23.39$, $\alpha=-1.09$, and $\phi_*=1.16\times10^{-2}\ {\rm Mpc}^{-3}$ \cite{Kochanek:2000im}, and convert them to the observed convention through
\begin{equation}
M_*^{\rm obs}=M_*+5\log_{10}h,
\qquad
\phi_*^{\rm obs}=\phi_* h^3.
\end{equation}

\begin{figure}[htbp]
    \centering
    \includegraphics[width=1\linewidth]{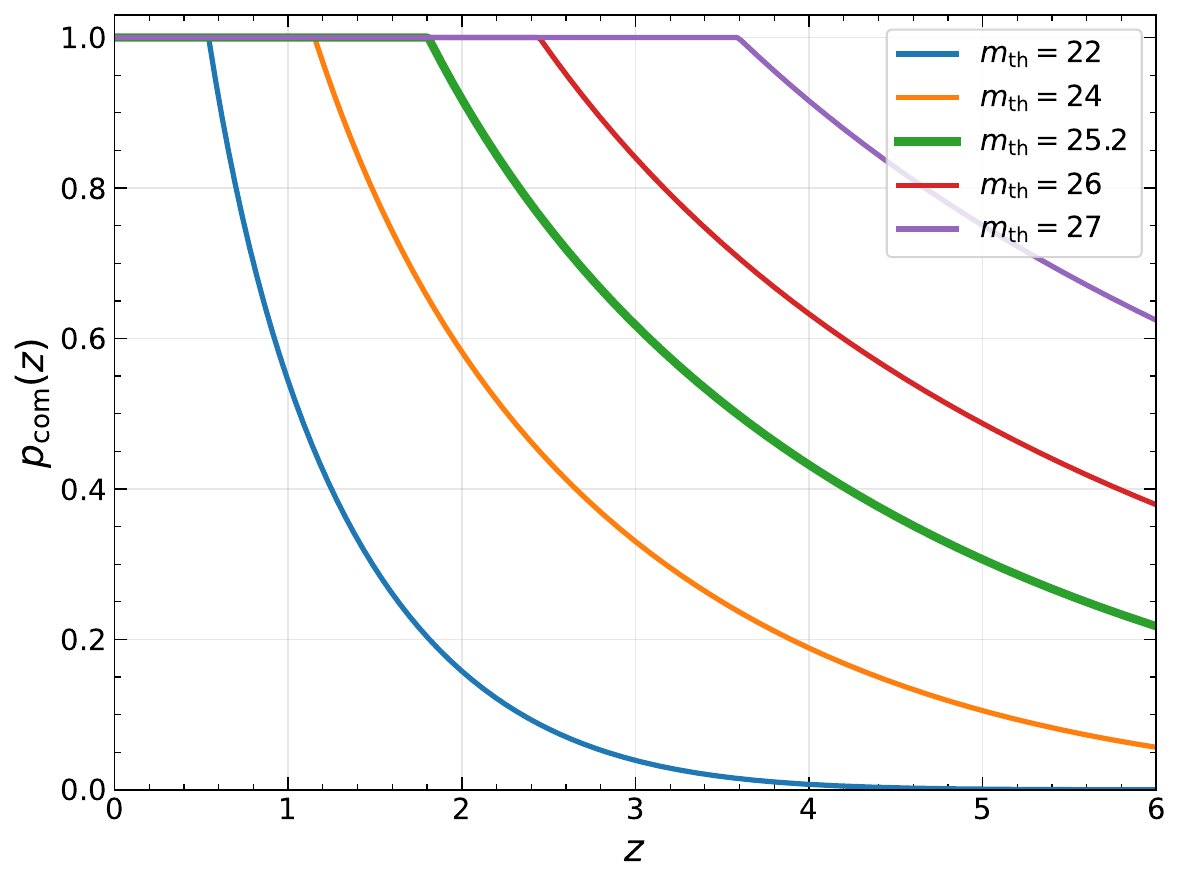}
    \caption{Catalog completeness $p_{\rm com}(z)$ versus redshift for several limiting apparent magnitudes $m_{\rm th}$, computed from the Schechter luminosity function of Ref.~\cite{Kochanek:2000im}. The thick green curve is the baseline $m_{\rm th}=25.2$; deeper surveys (larger $m_{\rm th}$) remain complete to higher redshift.}
    \label{fig:pcom}
\end{figure}

Given an apparent-magnitude threshold $m_{\rm th}$, the corresponding absolute-magnitude limit at redshift $z$ is
\begin{equation}
M_{\rm th}(z)=m_{\rm th}-\mu(z)-K(z),
\end{equation}
where $\mu(z)$ is the distance modulus. The $K$-correction is modeled as \cite{Kochanek:2000im}
\begin{equation}
K(z)=
\begin{cases}
-6\log_{10}(1+z), & z\le 0.3,\\
-6\log_{10}(1+0.3), & z>0.3.
\end{cases}
\end{equation}
The catalog completeness is then defined as
\begin{equation}
p_{\rm com}(z)
=
\frac{\int_{M_{\rm min}^{\rm obs}}^{M_{\rm th}(z)} dM\, \Phi(M)}
{\int_{M_{\rm min}^{\rm obs}}^{M_{\rm max}^{\rm obs}} dM\, \Phi(M)},
\end{equation}
where $M_{\rm min}^{\rm obs}$ and $M_{\rm max}^{\rm obs}$ denote the magnitude bounds corresponding to $M_{\rm min}=-27$ and $M_{\rm max}=-19$ in the source-frame convention. The complementary fraction,
\begin{equation}
p_{\rm miss}(z)=1-p_{\rm com}(z),
\end{equation}
accounts for galaxies missed by the magnitude-limited catalog. Fig.~\ref{fig:pcom} shows $p_{\rm com}(z)$ for several limiting magnitudes; for the baseline $m_{\rm th}=25.2$ the catalog is essentially complete out to $z\simeq1.5$ and falls to $\simeq0.6$ by $z=3$. {This depth is comparable to that of forthcoming deep optical surveys such as LSST~\cite{Ivezic:2019lsst} and CSST~\cite{Gong:2019yxt,Qu:2023abc}, and the resulting completeness is consistent with expectations for such catalogs.}

\subsubsection{Construction of candidate host galaxies}

For the $i$-th GW event, the expected number of galaxies contained in the observed catalog is
\begin{equation}
N_{{\rm exp},i}
=
\mathrm{round}\!\left[
\Delta\Omega_{i} \,\phi_*^{\rm obs}
\int dz\,
\frac{dV_c}{dz\,d\Omega}(z)\,
p_{\rm com}(z)
\right],
\end{equation}
where $\Delta\Omega_{i}$ is the GW sky-localization area.

Candidate galaxies are drawn from the completeness-weighted comoving-volume distribution,
\begin{equation}
p_{\rm draw}(z)\propto \frac{dV_c}{dz}(z)\,p_{\rm com}(z),
\end{equation}
with angular positions sampled uniformly over the GW localization region. Each candidate galaxy $j$ is assigned an angular weight according to its consistency with the GW sky localization,
\begin{equation}
w_{ij}\propto
\exp\!\left[
-\frac{1}{2}
(\mathbf{x}_{ij}-\mathbf{x}^{\rm inj}_i)^{\rm T}
\Sigma_{{\rm sky},i}^{-1}
(\mathbf{x}_{ij}-\mathbf{x}^{\rm inj}_i)
\right],
\end{equation}
where $\mathbf{x}_{ij}=(\alpha_{ij},\delta_{ij})$, $\mathbf{x}^{\rm inj}_i$ is the injected sky position of the $i$-th GW event, and $\Sigma_{\rm sky}$ is the $2\times 2$ covariance matrix in $(\alpha,\delta)$ obtained from FIM, as shown in Eq.~\eqref{eq:sky_localization}. The weights are normalized such that
\begin{equation}
\sum_{j=1}^{N_{{\rm exp},i}} w_{ij}=1.
\end{equation}

The redshift distribution of cataloged candidate hosts is modeled as a weighted Gaussian mixture,
\begin{equation}
p_{{\rm cat},i}(z)
=
\sum_{j=1}^{N_{{\rm exp},i}}
w_{ij}\,
\mathcal{N}\!\left[z; z_{ij}, \sigma_z(1+z_{ij})\right],
\end{equation}
where $\sigma_z(1+z)$ is the galaxy redshift uncertainty. With the above normalization of $w_{ij}$, $p_{{\rm cat},i}(z)$ is normalized to unity.

Eq.~\eqref{eq:Lambda_i} consistently incorporates both the observed and unobserved galaxy populations: the first term describes the contribution from cataloged galaxies conditioned on the realized mock catalog, while the second statistically restores potential hosts below the magnitude threshold.

\subsubsection{Selection effects}

Selection effects are incorporated through the probability for an event to enter the analysis sample as a function of luminosity distance. We estimate this quantity from precomputed Monte Carlo realizations of the network SNR on a luminosity-distance grid. At each grid point, the selection probability is defined as the fraction of realizations satisfying the analysis threshold,
\begin{equation}
p_{\rm det}(d_{\rm L})
=
\frac{1}{N_{\rm samp}}
\sum_{s=1}^{N_{\rm samp}}
\mathcal{H}(\rho_{s}-\rho_{\rm th}),
\end{equation}
where $\mathcal{H}$ is the Heaviside step function. We use $N_{\rm samp}=5000$ Monte Carlo realizations at each grid point and tabulate the selection function over 3000 luminosity-distance samples. In this construction, $p_{\rm det}(d_{\rm L})$ is obtained by averaging over the remaining source and geometric parameters at fixed luminosity distance. The resulting $p_{\rm det}(d_{\rm L})$ is interpolated as a continuous function of luminosity distance and used in Eq.~\eqref{eq:likelihood} to account for selection bias. We note that the network SNR entering this selection function is evaluated with the redshifted, detector-frame chirp mass $\mathcal{M}_c(1+z)$ at the redshift corresponding to each luminosity distance through the fiducial $d_{\rm L}$--$z$ relation; the dependence of the detection probability on redshift, via the redshifted chirp mass, is therefore fully included, and tabulating $p_{\rm det}$ as a function of $d_{\rm L}$ is equivalent to tabulating it as a function of $z$.

\subsection{Cosmological models and conventional late-universe probes}

In this work, we consider two representative cosmological models: $\Lambda$CDM and $w$CDM. In the $\Lambda$CDM model, we mainly discuss the constraint on $H_0$ and include the $w$CDM model to discuss the constraint on the dark energy EoS parameter.

In a Friedmann-Lema\^{\i}tre-Robertson-Walker universe, the comoving distance is
\begin{equation}
D_{\rm C}(z)=\int_0^z \frac{c\,{\rm d}z'}{H(z')},
\end{equation}
where $H(z)$ is the Hubble parameter, and the transverse comoving distance is given by \cite{Hogg:1999ad}
\begin{equation}
D_{\rm M}(z)=
\begin{cases}
\dfrac{c}{H_0\sqrt{\Omega_K}}
\sinh\!\left(\dfrac{H_0\sqrt{\Omega_K}}{c}D_{\rm C}(z)\right), & \Omega_K>0,\\[6pt]
D_{\rm C}(z), & \Omega_K=0,\\[6pt]
\dfrac{c}{H_0\sqrt{-\Omega_K}}
\sin\!\left(\dfrac{H_0\sqrt{-\Omega_K}}{c}D_{\rm C}(z)\right), & \Omega_K<0,
\end{cases}
\end{equation}
where $\Omega_K$ is the spatial curvature parameter, and the luminosity distance and Hubble distance are given by
\begin{equation}
D_{\rm L}(z)=(1+z)D_{\rm M}(z),\qquad D_{\rm H}(z)=\frac{c}{H(z)}.
\end{equation}
For the flat $\Lambda$CDM model here, the Hubble parameter is given by
\begin{equation}
H(z)=H_0\sqrt{\Omega_{\rm m}(1+z)^3+(1-\Omega_{\rm m})},
\end{equation}
where $\Omega_{\rm m}$ is the density parameter of matter. For the flat $w$CDM model, the dark energy EoS parameter $w$ can deviate from -1, and the Hubble parameter is
\begin{equation}
H(z)=H_0\left[\Omega_{\rm m}(1+z)^3+(1-\Omega_{\rm m})(1+z)^{3(1+w)}\right]^{1/2}.
\end{equation}

When probing dark energy, a single cosmological probe often suffers from strong parameter degeneracies. We therefore complement the GW data with two mainstream late-universe probes: baryon acoustic oscillations (BAOs) and Type Ia supernovae (SNe Ia). For BAO, we use the DESI DR2 measurements \cite{DESI:2025zgx}, including $D_{\rm M}(z)/r_{\rm d}$, $D_{\rm H}(z)/r_{\rm d}$, and $D_{\rm V}(z)/r_{\rm d}\equiv \bigl[zD_{\rm M}^2(z)D_{\rm H}(z)\bigr]^{1/3}$. $r_{\rm d}$ is the sound horizon at the baryon drag epoch. To avoid imposing early-universe information, we treat $r_{\rm d}$ as a free parameter. For SNe Ia, we use the DESY5 sample \cite{DES:2024upw,DES:2024jxu}. The observed peak magnitude is related to the luminosity distance through
\begin{equation}
m_B(z)=M_B+5\log_{10}\!\left[\frac{D_{\rm L}(z)}{\rm Mpc}\right]+25,
\end{equation}
where the absolute magnitude $M_B$ is also treated as a free parameter.

For both BAO and SNe Ia, the likelihood is given by
\begin{equation}
\mathcal{L}\propto
\exp\!\left[
-\frac{1}{2}
(\boldsymbol{X}_{\rm th}-\boldsymbol{X}_{\rm obs})^{\rm T}
\boldsymbol{C}^{-1}
(\boldsymbol{X}_{\rm th}-\boldsymbol{X}_{\rm obs})
\right],
\end{equation}
where $\boldsymbol{C}$ is the covariance matrix, and $\boldsymbol{X}_{\rm th}$ and $\boldsymbol{X}_{\rm obs}$ denote the theoretical and observed data vectors, respectively. For BAO, $\boldsymbol{X}$ consists of the measured distance ratios, while for SNe Ia it corresponds to the apparent-magnitude measurements. The data and likelihood codes for DESI DR2 BAO\footnote{\url{https://github.com/CobayaSampler/bao_data/tree/master}} and DESY5 SNe Ia\footnote{\url{https://github.com/des-science/DES-SN5YR}} are publicly available.


\section{Results and discussion}\label{sec:results}

\subsection{\texorpdfstring{$H_0$ constraints under three-band networks: comparison among detector configurations}{H0 constraints under three-band networks: comparison among detector configurations}}

In this subsection, all comparisons are performed under a common fiducial setup. Specifically, we adopt the baseline IMBHB population model, $z2$, an observation duration of $4~{\rm yr}$, a GW event selection threshold of $\rho>100$, a mock galaxy catalog with limiting magnitude $m_{\rm th}=25.2$, and a galaxy redshift uncertainty of $\sigma_z=0.02$. Under this default setup, in the $\Lambda$CDM model, we compare constraints on $H_0$ and $\Omega_{\rm m}$ from LGWA+TJ, LGWA+ET, TJ+ET, and LGWA+TJ+ET, respectively, summarized in Table~\ref{tab:H0}.

\begin{figure}[htbp]
\centering
\includegraphics[width=1\linewidth]{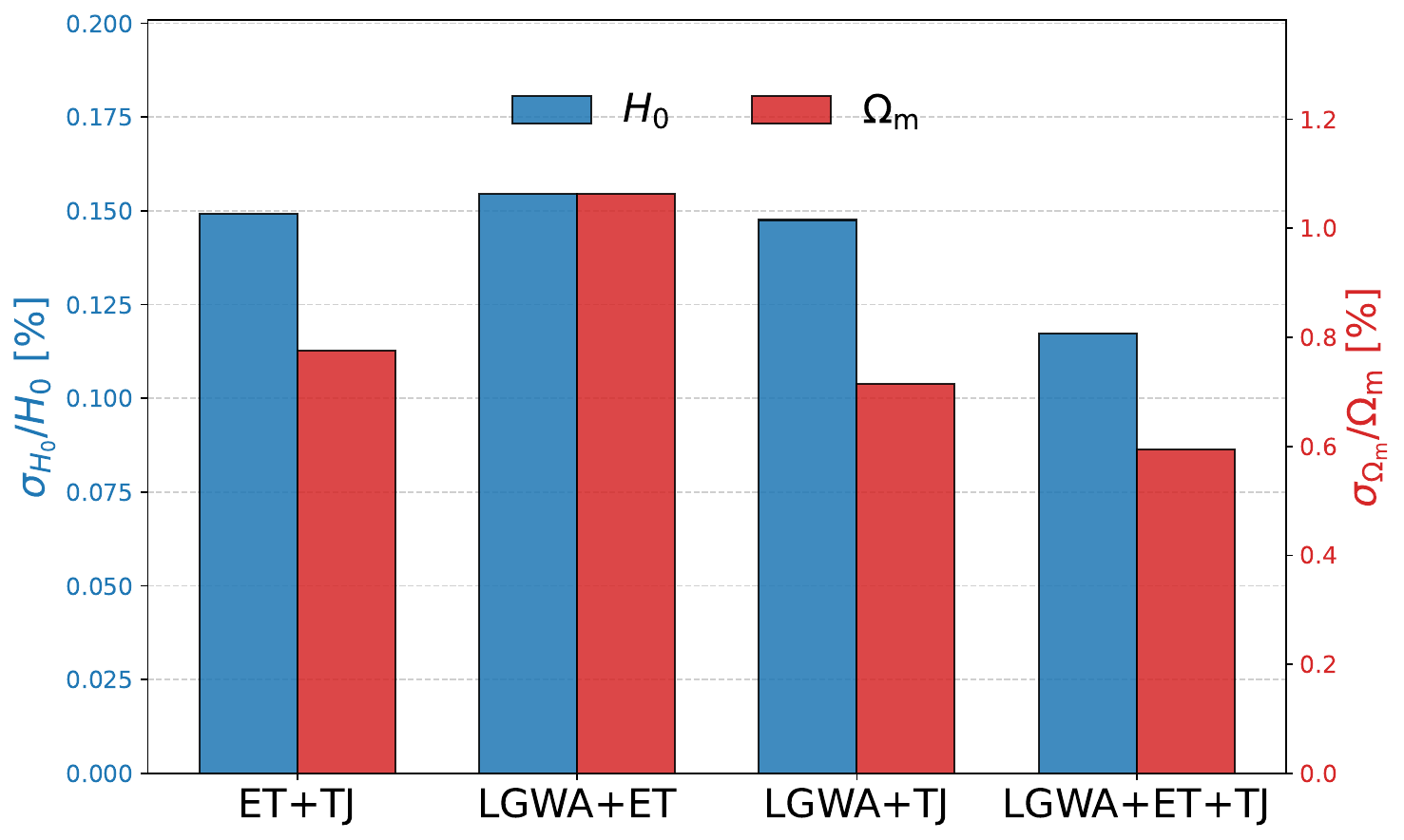}
\caption{Constraint precisions of $H_0$ and $\Omega_{\rm m}$ for different detector network configurations. For each network configuration, the blue bar (referencing the left y-axis) and the red bar (referencing the right y-axis) represent the relative uncertainties $\sigma_{H_0}/H_0$ and $\sigma_{\Omega_{m}}/\Omega_{m}$ in percentage, respectively.}
\label{fig:H0_networks}
\end{figure}

Fig.~\ref{fig:H0_networks} shows that the cosmological constraints depend on the detector-network configuration. For $H_0$, the forecasted precision is already at the sub-percent level for all networks considered here. The full LGWA+ET+TJ network yields the tightest constraint, with $\sigma(H_0)/H_0 \sim 0.12\%$. This is about $20\%$ tighter than the $\sim 0.15\%$ constraints from two-detector networks, demonstrating the advantage of the three-band GW network in constraining $H_0$. For $\Omega_{\rm m}$, the constraint precisions exhibit a considerably stronger dependence on the network configuration. The fractional uncertainty decreases from $\sim 1.1\%$ for LGWA+ET to $\sim 0.6\%$ for LGWA+ET+TJ, corresponding to an improvement of nearly a factor of two. This behavior suggests that $\Omega_{\rm m}$ is more sensitive than $H_0$ to the overall redshift reach and to the reduction of measurement errors in $d_{\rm L}$ enabled by a broader multiband network. It is also noteworthy that the ET+TJ and LGWA+TJ networks yield comparable cosmological constraints, both of which outperform LGWA+ET. This ranking may appear surprising, since, as shown in Fig.~\ref{fig:CDF}, the per-event $d_{\rm L}$ and $\Delta\Omega_{90\%}$ precisions of LGWA+ET are not the worst among the four configurations. However, as indicated by the detection yields in Fig.~\ref{fig:Nz}, LGWA+ET produces the fewest usable events, which ultimately limits its cosmological constraining power. Overall, the results in Fig.~\ref{fig:H0_networks} demonstrate that the joint LGWA+ET+TJ configuration delivers the most stringent constraints on both parameters, underscoring the advantage of combining complementary frequency bands for dark-siren cosmology.

\begin{table}
\renewcommand\arraystretch{1}
\caption{Relative uncertainties of $H_0$ and $\Omega_{\rm m}$ for different detector network configurations under the baseline setup ($z2$ population model, 4-year observation, $\rho>100$, $m_{\rm th}=25.2$, $\sigma_z=0.02$).}
\label{tab:H0}
\centering
\begin{tabular}{ccc}
\hline\hline
\makebox[0.15\textwidth][c]{Network} & \makebox[0.15\textwidth][c]{$\sigma_{H_0}/H_0$} & \makebox[0.15\textwidth][c]{$\sigma_{\Omega_{\rm m}}/\Omega_{\rm m}$}\\
\hline
ET+TJ         &  0.149\%        &   0.775\%          \\
LGWA+ET         &  0.155\%        &   1.063\%          \\
LGWA+TJ         &  0.148\%       &    0.714\%        \\
LGWA+ET+TJ         &  0.117\%        &   0.594\%          \\
\hline
\hline
\end{tabular}
\end{table}

\subsection{Dark-energy constraints from the combination of GW dark sirens, BAO, and SNe Ia}

Probing dark energy benefits from a purely late-universe observational framework because dark energy dominates the cosmic expansion at late times. Combining GW dark sirens with BAO and SNe~Ia thus provides a fully late-universe route to constrain both $H_0$ and the dark-energy EoS simultaneously \cite{Song:2025bio}.

\begin{figure}[htbp]
    \centering
    \includegraphics[width=1\linewidth]{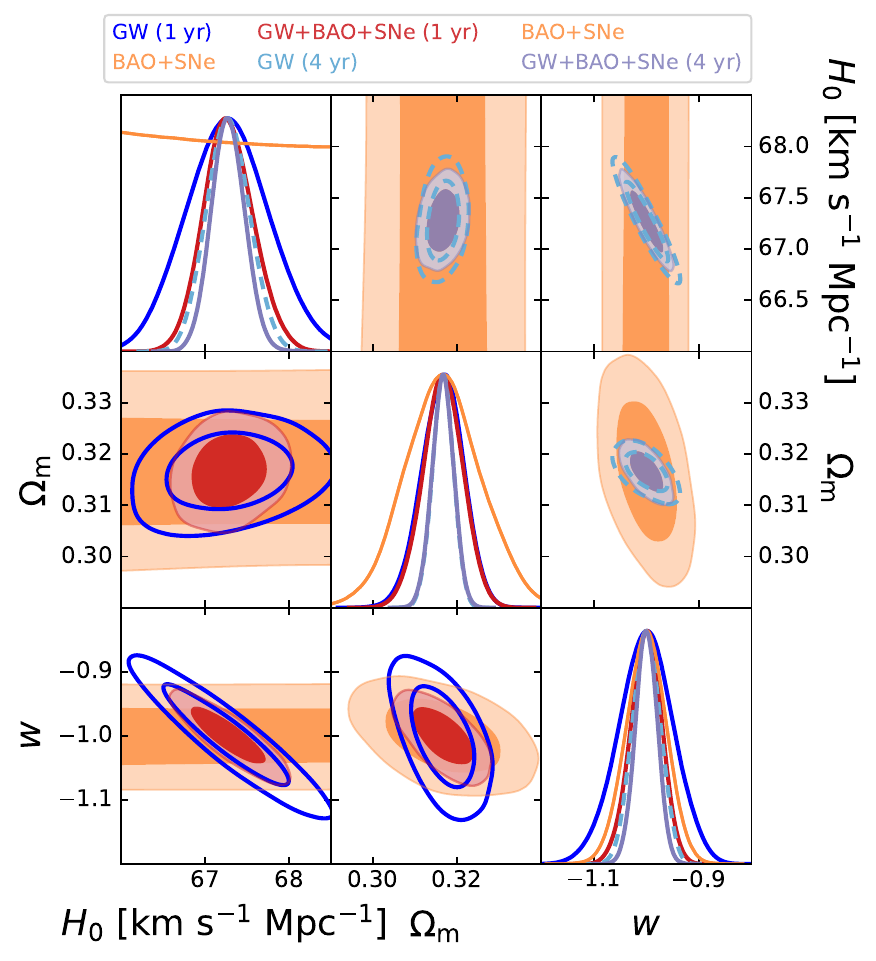}
    \caption{Two-dimensional posterior distributions in the $w$CDM model for different combinations of GW, BAO, and SNe Ia data. The lower-left triangle shows the constraints from the 1-year GW sample, while the upper-right triangle shows those from the 4-year GW sample.}
    \label{fig:w0}
\end{figure}

We work within the $w$CDM model and consider GW, BAO+SNe~Ia, and GW+BAO+SNe~Ia datasets for the representative LGWA+TJ+ET configuration. As in the previous analysis, we adopt the fiducial $z2$ population model and the detection threshold $\rho>100$. The parameter set $\{H_0,\,\Omega_{\rm m},\,w,\,r_{\rm d},\,M_B\}$ is allowed to vary freely, so that the resulting constraints are independent of both CMB information and distance-ladder calibration. Fig.~\ref{fig:w0} shows the posterior distributions in the $(H_0,\Omega_{\rm m},w)$ space, with the lower-left triangle corresponding to the 1-year GW sample and the upper-right triangle to the 4-year sample. To facilitate comparison among different datasets, we shift the posterior distributions in the figure so that their medians align at $H_0=67.27$, $\Omega_{\rm m}=0.3166$, and $w=-1$. Table~\ref{tab:w} summarizes the constraint precisions. Focusing on $w$, the relative uncertainty from GW dark sirens alone improves from $5.132\%$ (1 year) to $2.656\%$ (4 years), showing that increasing the GW sample size is highly effective for dark-energy inference. For the 1-year case, adding BAO+SNe~Ia tightens the $w$ constraint from $5.132\%$ to $2.947\%$, corresponding to an improvement of about $42.6\%$. For the 4-year case, the improvement from adding BAO+SNe~Ia is more modest, from $2.656\%$ to $2.069\%$ (about $22.1\%$). Notably, the 4-year GW constraint on $w$ is already tighter than the BAO+SNe~Ia result, and the GW (4 yr)+BAO+SNe~Ia combination achieves constraints comparable to, and slightly tighter than, those from the current CMB+BAO+SNe~Ia combination \cite{DESI:2025zgx}. The BAO and SNe~Ia likelihoods adopted here are based on existing observations. Future dark-energy surveys such as Euclid~\cite{Euclid:2024yrr}, LSST~\cite{LSSTDarkEnergyScience:2018yem}, and CSST~\cite{CSST:2025ssq} will provide substantially more precise BAO and SNe measurements and should further tighten the joint constraints.

\begin{table}
\renewcommand\arraystretch{1}
\caption{Relative uncertainties of $H_0$, $\Omega_{\rm m}$, and $w$ in the $w$CDM model for different data combinations using the LGWA+TJ+ET network under the baseline setup ($z2$ population model, $\rho>100$, $m_{\rm th}=25.2$, $\sigma_z=0.02$).}
\label{tab:w}
\centering
\begin{tabular}{cccc}
\hline\hline
\makebox[0.07\textwidth][c]{Data} & \makebox[0.07\textwidth][c]{$\sigma_{H_0}/H_0$} & \makebox[0.07\textwidth][c]{$\sigma_{\Omega_{\rm m}}/\Omega_{\rm m}$} & \makebox[0.07\textwidth][c]{$\sigma_{w}/|w|$}\\
\hline
GW (1 yr)                    &  0.735\%        &   1.550\%  &   5.132\%        \\
GW (4 yr)                    &  0.376\%        &   0.782\%  &   2.656\%     \\
BAO+SNe Ia                   &  $\sim$         &   2.868\%  &   3.807\%  \\
GW (1 yr) + BAO + SNe Ia   &  0.435\%        &   1.489\%  &   2.947\%     \\
GW (4 yr) + BAO + SNe Ia   &  0.296\%        &   0.820\%  &   2.069\%     \\
\hline
\hline
\end{tabular}
\end{table}

\subsection{Influence of population-model and galaxy-catalog assumptions}

\begin{figure}[htbp]
    \centering
    \includegraphics[width=\linewidth]{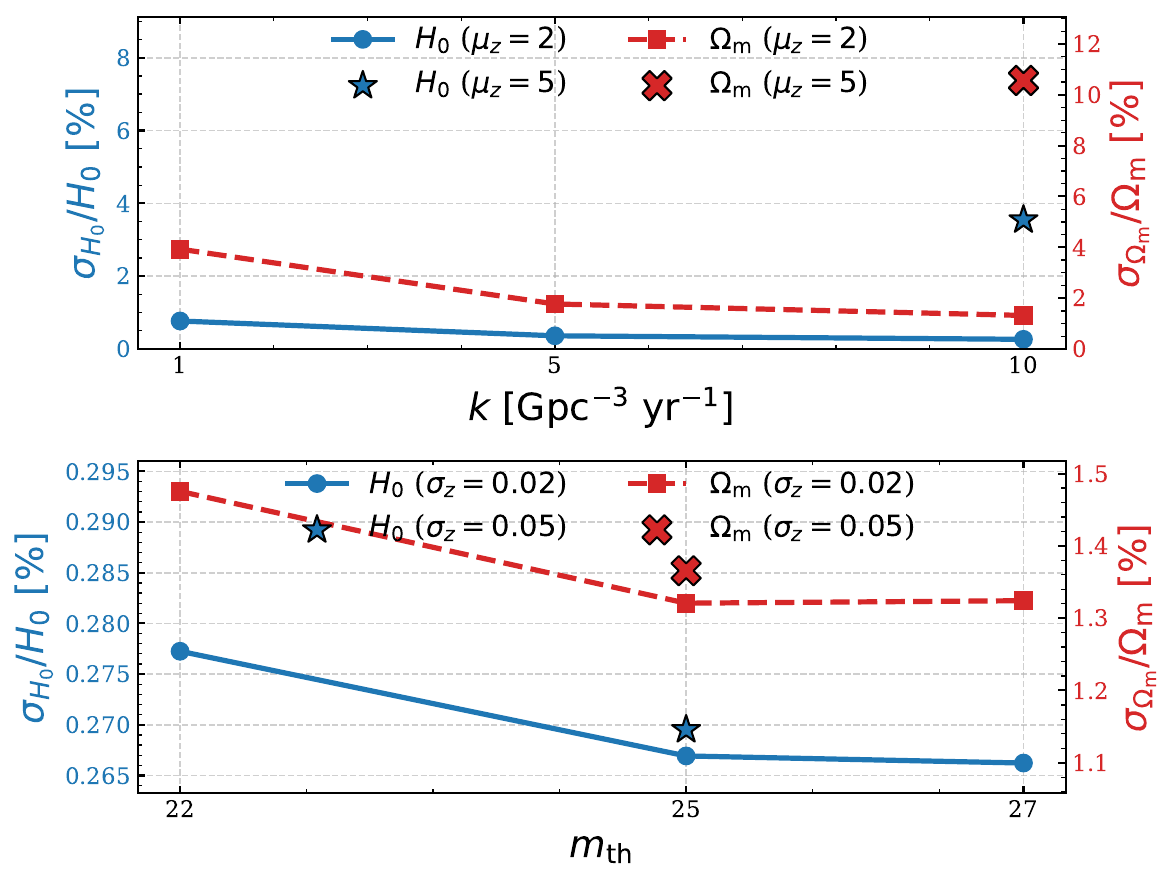}
    \caption{The influence of population-model and galaxy-catalog assumptions on the constraint precisions of $H_0$ and $\Omega_{\rm m}$. In both panels, the blue solid lines with circles (referencing the left y-axis) and the red dashed lines with squares (referencing the right y-axis) represent the relative uncertainties of $H_0$ and $\Omega_{\rm m}$, respectively, in 1-year observational duration. The top panel shows the constraint precisions as a function of the merger rate parameter $k$ under the population assumption with a redshift peak at $\mu_z=2$. The scattered blue star and red cross at $k=10$ represent the results for an alternative population model with $\mu_z=5$. The bottom panel shows the precisions versus the apparent magnitude threshold $m_{\rm th}$ of the galaxy catalog, assuming a default photometric redshift error of $\sigma_z=0.02$. The scattered star and cross at $m_{\rm th}=25$ demonstrate the corresponding precisions when a larger redshift uncertainty of $\sigma_z=0.05$ is assumed.}
    \label{fig:pop_cat}
\end{figure}

Uncertainties in the IMBHB population model and observational limitations of galaxy catalogs, especially survey incompleteness and photometric redshift errors, constitute two major sources of systematic uncertainties in dark-siren cosmology. To quantify how they affect the inferred constraints on $H_0$ and $\Omega_{\rm m}$, we perform a set of controlled forecast tests. We fix the detector network to LGWA+TJ+ET, adopt a 1-year observing time, and vary the GW source-population model and the galaxy-catalog parameters separately. Fig.~\ref{fig:pop_cat} shows the resulting constraint precisions under different population and catalog assumptions.

The top panel of Fig.~\ref{fig:pop_cat} shows how source-population models affect cosmological constraints. As expected, a larger merger-rate parameter $k$ monotonically improves the constraints on both $H_0$ and $\Omega_{\rm m}$, mainly by increasing the number of detectable events. For reference, at the baseline $k=10$ the one-year LGWA+ET+TJ sample used here contains $N_{\rm GW}=12{,}805$ events with $\rho>100$ (and $50{,}936$ for four years); the curves therefore quantify how the constraint precision depends on the population model and the corresponding number of events, and indicate directly how the forecasts would change for any other assumed merger rate. The redshift distribution of mergers also plays an important role. To illustrate this effect, we compare the fiducial model, which peaks at $\mu_z=2$, with an alternative model that favors higher-redshift mergers, with $\mu_z=5$, at fixed $k=10$. The isolated markers (the blue star and red cross) show that the $\mu_z=5$ model yields weaker constraints. Fig.~\ref{fig:Nz} clarifies this trend. In the $z5$ model with $\mu_z=5$, both the event rate and the detection rate decrease, which directly reduces the statistical power of the sample. Events at higher redshifts also tend to have lower SNRs, leading to poorer distance measurements and sky localization. In addition, the galaxy catalog becomes less complete and provides worse redshift measurements at higher redshifts. Together, these effects degrade the cosmological constraints.

The bottom panel of Fig.~\ref{fig:pop_cat} shows the effects of galaxy-catalog systematics, in particular catalog completeness and redshift-measurement errors. We vary the limiting apparent magnitude $m_{\rm th}$ from 22 to 27 and adopt a baseline photometric-redshift error of $\sigma_z=0.02$. A deeper survey, corresponding to a larger $m_{\rm th}$, yields a more complete catalog and increases the chance that the true GW host appears among the candidate galaxies. As a result, the relative uncertainties in $H_0$ and $\Omega_{\rm m}$ decrease as the catalog becomes deeper and completeness improves. Photometric-redshift precision also plays an important role. At fixed $m_{\rm th}=25$, increasing the redshift uncertainty from $\sigma_z=0.02$ to $\sigma_z=0.05$ (marked by the isolated symbols) noticeably weakens the cosmological constraints.

\subsection{Caveats and limitations}

Our forecasts are subject to several caveats, which we now make explicit.

\textit{Calibration and waveform systematics.} Our forecast assumes ideal detector calibration and a single waveform family. Calibration uncertainties in the strain amplitude and phase propagate into the inferred luminosity distances, and hence into the cosmological parameters; for current detectors they are at the percent level~\cite{Sun:2020wke}. More importantly, the high SNRs of three-band IMBHB events make the statistical errors so small that even minor waveform inaccuracies can bias the recovered parameters beyond the statistical level~\cite{Cutler:2007mi,Purrer:2019jcp}. Realizing the forecast precision therefore requires both effects to be controlled below the statistical floor.

\textit{Spin, precession, and higher modes.} We adopt the IMRPhenomD waveform~\cite{Husa:2015iqa,Khan:2015jqa}, which models only the dominant $(2,2)$ mode of aligned-spin, non-precessing binaries, thereby neglecting precession and higher-order modes that would otherwise help break the degeneracy between luminosity distance and inclination~\cite{Yun:2023ygz}. Furthermore, the component spins are fixed to zero rather than varied, so our nine-dimensional Fisher matrix omits the spin--distance correlations. Both simplifications render our distance, and hence cosmological, uncertainties somewhat optimistic; a full treatment including spins and a precessing, higher-mode waveform would broaden the posteriors.

\textit{Fisher-matrix approximation.} All uncertainties quoted here are computed with the Fisher information matrix, a linearized, high-SNR approximation that saturates the Cram\'er--Rao bound and is exact only for large SNR and Gaussian noise~\cite{Cutler:1994ys,Vallisneri:2007ev}. At finite SNR the true posteriors are generally broader and may be non-Gaussian, so our Fisher errors represent the most optimistic attainable precision, and a full Bayesian analysis would yield somewhat weaker constraints. The high SNRs of three-band IMBHB events make the approximation comparatively reliable here, but the quoted precisions remain idealized lower bounds.

\textit{Dataset combination and the Hubble tension.} Our joint analysis combines the GW, BAO, and SNe~Ia likelihoods under the assumption that the three probes share a common cosmology, and to isolate the forecast precision we center all posteriors on the same fiducial values, aligning their medians at $H_0=67.27$, $\Omega_{\rm m}=0.3166$, and $w=-1$ (Fig.~\ref{fig:w0}). Our results therefore quantify only the attainable error bars and presuppose no particular central value or dataset preference, so this consistency issue does not affect them. In practice, however, the situation could differ: if the GW dark sirens alone favored a high $H_0$ consistent with the local distance-ladder (SH0ES) value~\cite{Riess:2021jrx} while CMB+BAO+SNe~Ia favored the lower value~\cite{Planck:2018vyg,DESI:2024mwx,DESI:2025zgx}, the datasets would be in tension~\cite{Verde:2019ivm} and their likelihoods could not be combined naively. The proper strategy is then to first assess their mutual consistency and to combine only concordant probes; an independent, late-universe dark-siren measurement of $H_0$ would itself provide a valuable arbiter of this tension~\cite{Chen:2017rfc,DES:2019ccw,Song:2025bio}. We defer a detailed treatment to future work.

\section{Conclusion}

Gravitational-wave standard sirens provide a unique late-universe probe for the Hubble constant and the dark energy EoS parameter. Most GW events are dark sirens, and their cosmological constraining power depends critically on precise sky localizations and luminosity-distance measurements. Multiband GW observations can significantly improve both and increase the GW detection rate. Among potential multiband sources, IMBHBs are especially promising because their signals span the millihertz, decihertz, and hectohertz bands, making them ideal targets for three-band observations with future GW detector networks.

In this work, we investigate the cosmological potential of IMBHB dark sirens observed by a three-band GW network composed of TJ, LGWA, and ET. We simulate detectable IMBHB populations, quantify the improvements in luminosity-distance measurements and sky localizations enabled by three-band observations, and propagate them into cosmological parameter constraints. Among all detector configurations considered here, the TJ-LGWA-ET network delivers the best performance. In the $\Lambda$CDM model, it gives the tightest constraints on both $H_0$, with $\sigma_{H_0}/H_0\sim0.12\%$, and $\Omega_{\rm m}$, with $\sigma_{\Omega_{\rm m}}/\Omega_{\rm m}\sim0.6\%$. Relative to two-detector networks, it improves the $H_0$ constraint by about 21$\%$--24$\%$ and the $\Omega_{\rm m}$ constraint by about 17$\%$--44$\%$. In the $w$CDM model, a 4-year GW sample already constrains $w$ to about 2.7\%, and the addition of BAO and SNe~Ia achieves a tighter $w$ constraint than the CMB+BAO+SNe Ia result.

These results lead to three conclusions. First, the TJ-LGWA-ET network significantly enhances the cosmological power of IMBHB dark sirens and highlights the value of continuous three-band coverage for late-universe cosmology. Second, this network provides a promising late-universe probe of dark energy, especially when combined with BAO and SNe~Ia, without relying on CMB calibration or the distance ladder. Third, both astrophysical and observational systematics affect the final constraints. The assumed IMBHB population, especially the merger-rate normalization and redshift peak, changes the event rate and thus alters the cosmological constraints. Galaxy-catalog completeness and galaxy redshift errors also play critical roles. To fully realize the cosmological potential of the TJ-LGWA-ET network, future analyses will require deeper galaxy surveys and more precise redshift measurements.

\begin{acknowledgments}

We thank Di Chen and Yu-Feng Zhou for providing the Taiji sensitivity curve used in this work.
This work was supported by the National Natural Science Foundation of China (Grants Nos.\ 12473001, 12575049, and 12533001), the National SKA Program of China (Grants Nos.\ 2022SKA0110200 and 2022SKA0110203), the China Manned Space Program (Grant No.\ CMS-CSST-2025-A02), the 111 Project (Grant No.\ B16009), and the China Scholarship Council.

\end{acknowledgments}

\bibliography{three_bands}

\end{document}